# Semiconductor-like photocarrier dynamics in Dirac Semimetal $Cd_3As_2$ film Probed with transient Terahertz Spectroscopy

Wenjie Zhang[1], Yunkun Yang[2], Peng Suo[1], Kaiwen Sun[1], Jun Peng[1], Xian Lin[1], Faxian Xiu[2], and Guohong Ma[1]

[1] Department of Physics, Shanghai University, Shanghai 200444, China

[2] State Key Laboratory of Surface Physics and Department of Physics, Fudan University, Shanghai 200433, China

The topological three-dimensional Dirac semimetal $Cd_3As_2$ has drawn great attention for the novel physics and promising applications in optoelectronic devices operating in the infrared and THz regimes. Among the extensive studies in the past decades, one intriguing debate is the underlined mechanism that governing the nonequilibrium carrier dynamics following photoexcitation. In this study, the temperature dependent photocarrier dynamics in $Cd_3As_2$ film has been investigated with time-resolved terahertz spectroscopy. The experimental results demonstrate that photoexcitation results in abrupt increase in THz photoconductivity, and the subsequent relaxation shows a single exponential relaxation for various temperatures and pump fluences. The relaxation time increase from 4.7 ps at 5 K to 7.5 ps at 220 K, while the lifetime remains almost constant of ~7.5 ps with temperature above 220 K. A Rothwarf-Taylor model was employed to fit the temperature dependent relaxation time, and a narrow energy gap of ~35 meV is obtained, which occurs around the Dirac node. Our THz spectroscopy results demonstrate that the photocarrier relaxation in $Cd_3As_2$ shows a semiconductor-like behavior, rather than hot carrier scatterings in graphene and most of metals.

Three dimensional (3D) Dirac semimetals (DSMs), with its gapless feature and linear energy dispersion in the energy momentum diagram, have attracted enormous interest as a new family of topological quantum materials [1–5]. Owing to the massless Fermion energy dispersion, DSMs usually have high carrier mobility and a tunable chemical potential, resulting in potential applications in electronics and optoelectronics [6-10]. Conventionally, DSMs have large absorption coefficient and ultrafast response in infrared and terahertz (THz) region [11–13], and thus show appealing application in designing the advanced optoelectronic devices for far infrared radiation. As a member of 3D DSMs, $Cd_3As_2$ has been paid considerable attention due to its high stability and large linear energy-momentum space. In the past decades, massive studies have been carried out for exploring the equilibrium transportation in $Cd_3As_2$, which revealed that the DSM $Cd_3As_2$ shows high carrier mobility [14-16], large magnetoresistance [17–19], magneto-optical response [20], and the nonlinear optical response in THz regime [21,22], To gain more practical applications, understanding the nonequilibrium carriers dynamics in $Cd_3As_2$ is important for developing high-speed and broadband photodetectors or modulators benefiting from its high carrier mobility and conical Dirac bands [23,24].

Although lots of studies have been carried out on the photocarrier dynamics in $Cd_3As_2$ following across bandgap excitation, the underlined mechanism governing the photocarrier dynamics of $Cd_3As_2$ is still under debate. In general, two different models are proposed to discuss the photocarrier dynamics in $Cd_3As_2$ with photoexcitation depending on the intra- and inter-band scattering rate [25]. If the interband scattering is faster than that of intraband scattering, the resultant quasi-equilibrium can be described by a Fermi-Dirac distribution. Carrier-phonon coupling leads to the cooling of hot carriers, and photocarrier dynamics can be interpreted with two-temperature model (TTM), [26] and this model has been widely accounted for the photocarrier dynamics in metals. [26-28] On the other hand, if the intra-band scattering is much faster than that of inter-band scattering/recombination, the photoexcited electrons and holes could establish the separated Fermi distribution with electrons in conduction band and holes in valence band, respectively. This picture is similar to semiconductor materials, in which Rothwarf-Taylor (RT) model is applicable.

Terahertz (THz) radiation has very low photon energy of a few meV, which makes the THz spectroscopy be a sensitive tool for measuring the conductivity change of the free carrier, it is especially useful to probe the ultrafast photoconductivity (PC) around Fermi surface of materials. In this letter, we utilize the time resolved THz spectroscopy to investigate photocarrier dynamics of a 50-nm $Cd_3As_2$ film with various pump fluence and temperature. Our experimental results demonstrate that optical pump of 780 nm leads to the sharp increase in THz positive PC in $Cd_3As_2$ film, the subsequent relaxation of PC follows a single exponential decay. Interestingly, the decaying time $\tau$ of PC shows a complex temperature dependence, which increases from 4.7 ps at 5 K to 7.5 ps at 220 K, while magnitude of $\tau$ remains almost constant when temperature is above 220 K. The THz PC dispersion obtained at various temperatures, pump fluences and delay times can be well fitted with Drude model. These THz experimental results reveal that the increase in THz PC after photoexcitation mainly arises from the increase in carrier population, and the subsequent relaxation is dominated by the electron-hole (e-h) recombination, which suggests that photocarrier dynamics in $Cd_3As_2$ shows a semiconductor-like behavior, and the RT model is employed to fit the temperature dependent THz relaxation time. The fitting results indicate that narrow band gap with magnitude of 35 meV exists at the Dirac points for the 50-nm $Cd_3As_2$ thin film. The coexistence of the high-frequency phonons (HFP) generation and e-h recombination results in a phonon bottleneck effect, which slows down the photocarrier dynamics at high temperature. Our THz spectroscopy results demonstrate that the photocarrier relaxation in $Cd_3As_2$ shows a semiconductor-like behavior, rather than hot carrier scatterings in graphene and most of metals.

Optical pump and terahertz probe (OPTP) experiments in the transmission configuration were performed to explore the dynamics of photocarriers in $Cd_3As_2$ thin film. The optical pulses are delivered from a Ti: sapphire amplifier with 120 femtoseconds (fs) duration at central wavelength of 780 nm (1.59 eV) and a repetition rate of 1 kHz. The THz emitter and detector are based on a pair of (110)-oriented ZnTe crystals. The optical pump and THz probe pulse are collinearly polarized with a spot size of 6.5 and 2.0 mm on the surface of sample, respectively. All measurements were conducted in dry nitrogen atmosphere, and the samples were placed in a cryostat with temperature varying from 5 to 300 K.

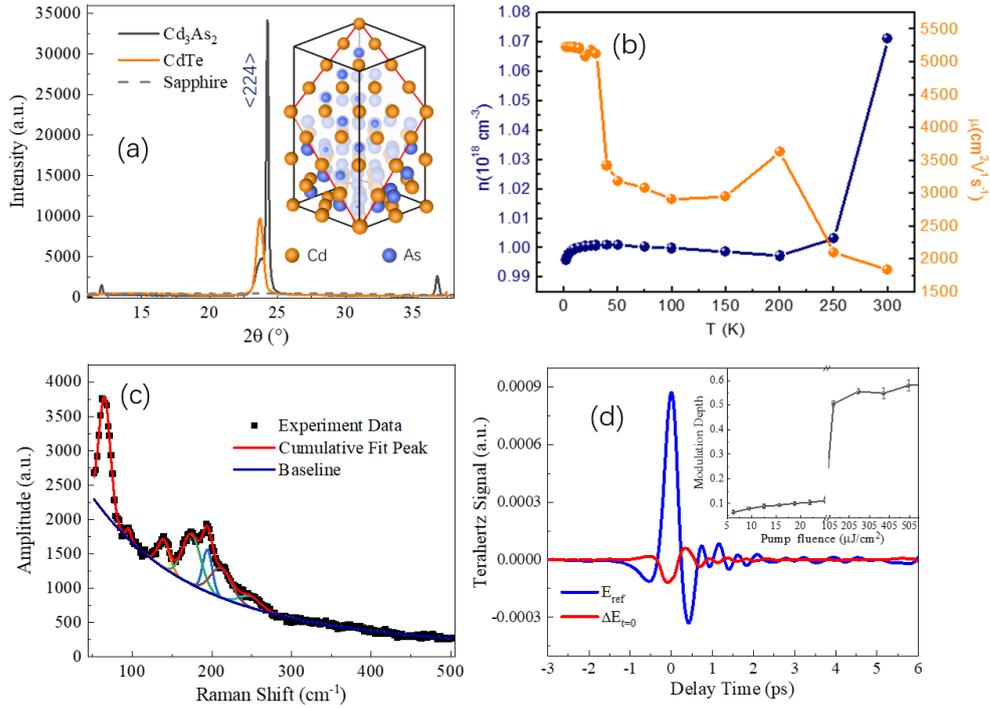

Fig 1(a) X-ray diffraction patterns of $Cd_3As_2$ film, the peak around $2\Theta=24.3^0$ corresponds to (224) diffraction. Inset illustrates the crystal structure of $Cd_3As_2$ along (112) plane. **(b) Temperature dependent carrier density and mobility of the 50 nm $Cd_3As_2$ film.** (c) Raman spectroscopy of $Cd_3As_2$ film along with Gaussian fittings. (d) THz waveform transmission of the sample without pump (blue) and with pump fluence of 18.75 μJ/cm² at delay time Δt = 0 ps, respectively. The inset shows the THz modulation with respect to pump fluence at room temperature.

The high quality $Cd_3As_2$ film were grown via a molecular beam epitaxy system on sapphire substrate. Before growing the $Cd_3As_2$ film, a 10 nm thick CdTe buffer layer was deposited to assist the $Cd_3As_2$ nucleation. The $Cd_3As_2$ film was grown epitaxially onto the buffer layer at 100°C, and the thickness is approximately 50 nm. The Fermi level of the intrinsic film is about 210 meV [29]. Figure 1a shows the X-ray diffraction (XRD) pattern of the $Cd_3As_2$ film. The sharp diffraction peak located around $2\Theta=24.3^0$ is indexed as (224) [30]. A shoulder peak at the left side of (224) is contributed by the (111) index of buffer layer CdTe. It is clear that the full-width at half maximum of the (224) peak is less than 0.18°, verifying the high crystallinity of the $Cd_3As_2$ thin film. Figure 1(b) shows the temperature dependent carrier density measured by the Hall effect. It is seen that the carrier concentration increases slightly from n≈$9.95×10^{17}$ $cm^{-3}$ at 5 K to $1.07×10^{18}$ $cm^{-3}$ at room temperature. The Raman scattering under 532 nm excitation along with Gaussian fitting are shown in Fig. 1(c), four pronounced peaks located around 62.2 ($B_{1g}$), 139.5 ($A_{1g}$), 174.3 ($B_{1g}$) and 194.5 $cm^{-1}$ ($B_{1g}$) are consistent to the previous reports in literatures [31,32], suggesting centrosymmetric structure for our 50 nm thick $Cd_3As_2$ film.

Figure 1(d) shows the transmitted THz waveform through the film at Δt=0 ps (red) and without pump beam (blue), respectively. The photoinduced change in the THz electric field is expressed as ΔE(t, Δt)=E(t, Δt)-$E_{ref}$(t), where E(t, Δt) and $E_{ref}$(t) is the time-domain THz waveform at Δt with and without pump, respectively, and *t* is the electro-optical sampling delay time for the distinction from the pump-probe delay time Δt. It is clear that THz transmitted waveforms collected at Δt=0 ps show out-of-phase with that of $E_{ref}$ (t), indicating the pump pulse induced the reduction of the THz transmission (increase of THz PC). By varying the delay time between the 780 nm optical and the THz pulses, the transient THz transmission can be mapped out through measuring the photoinduced peak absorption. The inset of Fig. 1(d) plots the THz peak modulation M=|(T(Δt=0)-$T_0$)/$T_0$| with respect to pump fluence at room temperature, in which T(Δt=0) and $T_0$ stand for the THz peak transmission at zero delay time with pump and without pump, respectively. The THz peak transmission shows linear dependence of pump fluence in both low and high pump fluence regimes: in a low fluence regime of F<25 μJ/$cm^2$, the THz modulation depth is less than 15% with a linear slope of 2.46 $mJ^{-1}$, while in high fluence regime of F>0.1 mJ/$cm^2$ regime, the THz modulation depth is higher than 50% with slope of 0.16 $mJ^{-1}$. Here, we mainly focused our study on the low pump fluence regime with F≤25 μJ/$cm^2$.

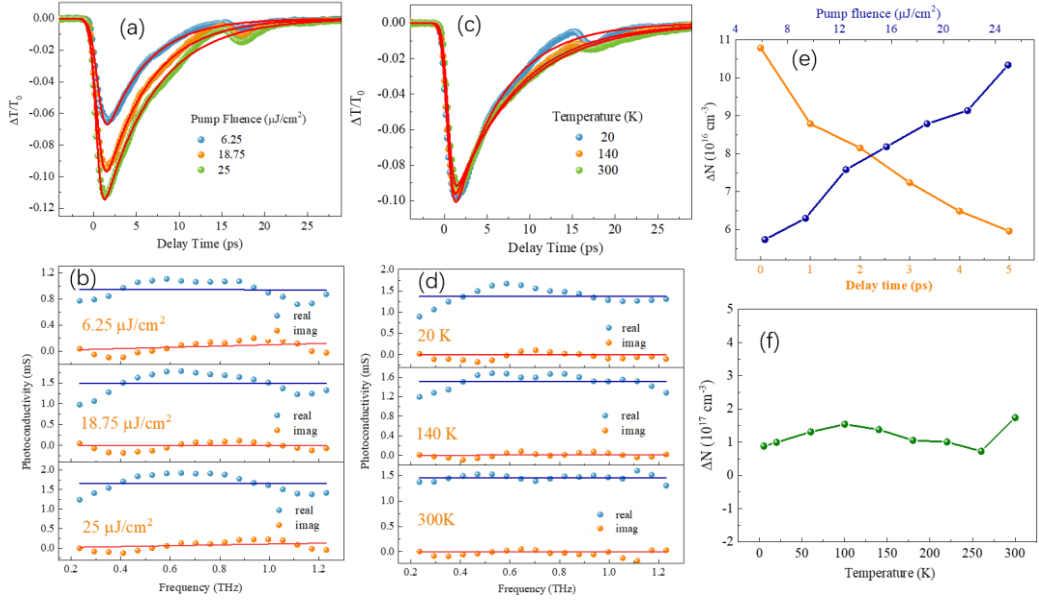

Figure 2 (a) Transient THz transmission under pump fluence of 6.25, 18.75 and 25 μJ/cm² at 5 K along with single exponential fitting (solid lines), (b) the corresponding PC dispersion measured at delay time Δt=0 in (a), with the solid lines of Drude model fitting. (c) Transient THz transmission at temperature of 20, 140 and 300 K under pump fluence of 18.75 μJ/cm², along with single exponential fitting (solid lines), (d) the corresponding PC dispersion obtained at delay time Δt=0 in (c), with the solid lines of Drude model fitting. (e) The fitting carrier concentration change (ΔN) with respect to pump fluence and delay time at 5 K. (f) The fitting carrier concentration change (ΔN) with respect to temperature obtained at **delay time of 1ps** and pump fluence of 18.75 μJ/cm².

Figures 2(a) and 2(c) show the transient THz transmission, ΔT/T₀ at 5 K with three selective pump fluences of 6.25, 18.75 and 25 μJ/cm² (a) as well as ΔT/T₀ with fixed pump fluence of 18.75 μJ/cm² for three selective temperatures of 20, 140 and 300 K (c). It is clear that the photoexcitation leads to the abrupt decrease in THz transmission, indicating the increase in THz photoconductivity for all measured temperatures and pump fluences. The transient THz trace can be well reproduced with single exponential function,

$$\frac{\Delta T}{T_0}(t) = A \cdot e^{\frac{\omega^2}{\tau^2} - \frac{t}{\tau}} \cdot erfc\left(\frac{2\omega}{\tau} - \frac{t}{2\omega}\right) + B \qquad (1)$$

in which, τ, A and B are the decay lifetime, amplitude and offset, respectively. 2ω=0.4 ps is the THz pulse width, and *efrc* (t)=1-erf (t) is a complementary error function. The best fittings are also presented with solid lines. The positive THz PC following photoexcitation suggests that the increase in the Drude weight is over than the increase in scattering rate with photoexcitation. The increase in Drude weight could be contributed by the elevated electron temperature and/or increase in carrier population. Due to the constraint of linear dispersion band structure in $Cd_3As_2$ semimetal, it could be possible that the increase in Drude weight is larger than that in scattering rate at elevated electron temperature caused by optical injection. [33,34] So that the free carriers' absorption of THz radiation is enhanced by hot electrons due to the intraband transition undergoes larger possible momentum and energy conservation spaces. Considering very large electron capacity ($\gamma_e \approx 70$ J $K^{-2}$ $m^{-3}$) in $Cd_3As_2$, [25,35] the increase in THz PC due to the elevated electron temperature is calculated to be less than 1% under pump fluence of 25 μJ/cm² (see S1 in supporting information). While the increase in THz PC is about 10%, which can be seen clearly from Figs. 2 (a) and (c), therefore, it is reasonable to assign the positive THz PC observed in Figs. 2 (a) and (c) to the increase of carrier population following photoexcitation.

In order to further verify the THz PC is contributed by the increase of photocarrier density ($\Delta N$) rather than increase in carrier temperature, we have also measured the PC at different pump fluence, delay time and temperatures, and the experimental results are displayed in Figs. 2 (b), (d), (e) and (f), respectively. For the 50-nm film of $Cd_3As_2$, Tinkham equation is applicable, in which transient THz transmission, time dependent $\Delta T/T_0$ can be transformed into complex PC in frequency region. According to the thin-film approximation, the real ($\Delta\sigma_r$) and imaginary ($\Delta\sigma_i$) PC is given by, [29, 36]

$$\Delta\sigma_r = \left(\frac{\cos\phi}{E} - 1\right)\frac{1+n_{sub}}{Z_0} \qquad (2a)$$

$$\Delta\sigma_i = -\frac{(1+n_{sub})\sin\phi}{EZ_0} \qquad (2b)$$

Where $E$ is the amplitude ratio of the Fourier transformed THz signal for unexcited sample and excited sample, and $\phi$ is the phase difference of these two, $n_{sub}$ and $Z_0$ are refractive index of substrate with $n_{sub}=3$ and free space impedance with $Z_0 \approx 377$ Ω, respectively. Figures 2(b) and 2(d) present the real and imaginary PC dispersion measured at 5 K with various pump fluences (2(b)), and different temperature with fixed fluence of 18.75 μJ/cm² (2(d)), both of them are measured at delay time of 1ps. The solid lines are the best-fitting with Drude model:

$$\Delta\tilde{\sigma} = \frac{\varepsilon_0 \omega_p^2 \tau_k}{1-i\omega\tau_k} = \frac{\Delta\sigma_{dc}}{1-i\omega\tau_k} \qquad (3)$$

in which, $\varepsilon_0$, $\omega_p$, and $\tau_k$ are the vacuum permittivity, plasma frequency and scattering time, respectively. Figure 2(e) presents the photocarrier density $\Delta N = \frac{\omega_p^2 \varepsilon_0 \tau_k}{e^2}$ with respect to pump fluence (blue) and delay time (orange), respectively. It is seen that ΔN increases with pump fluence and decreases with delay time, indicating photoexcitation does increase the carrier population, and the decay of photocarrier occurs via depopulation of carriers, probable via e-h recombination instead of hot carrier cooling. In addition, Fig. 2(f) presents the photocarrier density with respect to temperature, it is clearly seen that the photocarrier density almost does not change with temperature. The fitting momentum scattering time, $\tau_k$ with respect to pump fluence, delay time and temperature is given in S2 of supporting information, which shows good agreement with the prediction of photoexcitation induced the increase of carrier population.

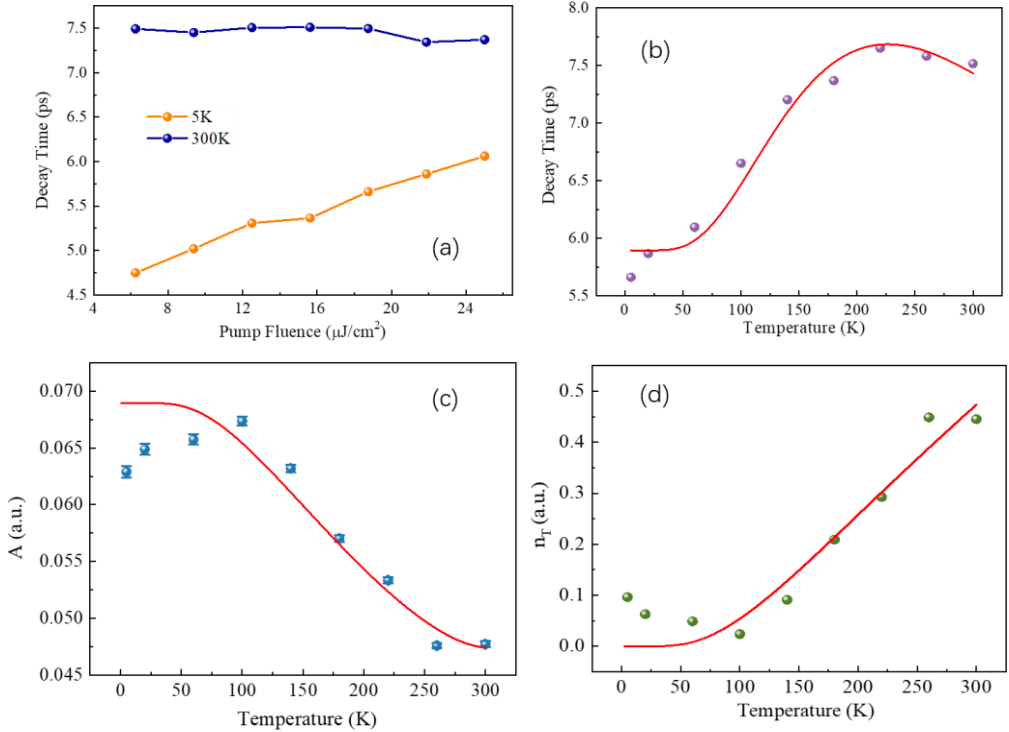

Figure 3 (a) The relaxation time τ obtained from single exponential fitting (Eq. (1)) with respect to pump feluence at 5 K (orange dots) and 300 K (blue dots). (b) The temperature dependent relaxation time τ (purple dots), the red line is the fitting curve with RT model. (c) The temperature dependent amplitude A (blue dots), the red line is the fitting curve with RT model. (d) Thermally populated carrier density per unit cell $n_T$ versus temperature calculated by RT model.

The THz experimental data presented in Fig. 2 clearly reveal that the positive PC of the intrinsic $Cd_3As_2$ film come from the increase of carrier population, which shows semiconductor-like behavior. The subsequent relaxation behaves both temperature and pump fluence dependence, which can be well reproduced with single exponential decay function. Obviously, the relaxation process is dominated by the photocarrier depopulation, so that we proposed that e-h recombination plays a dominated role in photocarrier relaxation in $Cd_3As_2$ following optical excitation. In the following part, we will discuss the possible originality of the photocarrier relaxation in $Cd_3As_2$ film. Figure 3 (a) presents the fitting time constants versus pump fluence at 5 and 300 K, respectively. It is obvious that, at low temperature of 5 K, the decay time slows down from 4.7 ps to 6.1 ps as the pump fluence increase from 6.5 µJ/cm$^2$ to 25 µJ/cm$^2$. While this change was not observed at room temperature, in which the lifetime remains 7.5 ps with various pump fluences. Figure 3(b) presents the fitting lifetime with respect to temperatures under pump fluence of 18.75 µJ/cm$^2$. Interestingly, it is noted that when the temperature is lower than 220 K, the relaxation time increases with temperature under identic pump fluence, and the relaxation time remains unchanged when the temperature is higher than 220 K.

The pump fluence dependence at room temperature (Fig. 3a) and temperature dependence of relaxation time (Fig. 3b) for our $Cd_3As_2$ thin film cannot be interpreted with TTM. According to the TTM, the electron-electron (e-e) thermalization is much faster than electron-phonon (e-ph) thermalization after photoexcitation, and the electrons with the elevated temperature transfer the excess energy to lattice *via* e-ph coupling until the two subsystems reach a balanced temperature. Therefore, the hot carrier relaxation time is expected to be longer as the pump fluence increase, which cannot interpret our experimental finding in Fig. 3 (a). Furthermore, the carrier lifetime remaining unchanged for temperature above 220 K is also beyond the TTM. It is noted that the e-ph coupling time is conventionally less than 2 ps in most of metal and semimetal materials, which is 3-fold faster than the relaxation time in our $Cd_3As_2$ film. This anomalous pump fluence and temperature dependence of relaxation time in the $Cd_3As_2$ film suggests different mechanism underlying dominates the photocarrier relaxation. It is noted that $Cd_3As_2$ will undergo semimetal to semiconductor transition with decreasing the film thickness [37], and gap opening in a $Cd_3As_2$ thin film has been predicted theoretically, and verified experimentally. Our previous study also indicated the semiconductor-like character for the photocarrier relaxation in $Cd_3As_2$ films. Therefore, it is reasonable to assume that our 50 nm thick film is thin enough to open a narrow gap at Dirac point, which will have a profound effect on the photocarrier dynamics following optical excitation. It has been demonstrated that RT model is applicable to analyze the nonequilibrium carrier dynamics in narrow gap structure, such as superconductor gap [38,39], charge density wave gap [40] as well as collective hybridization [41] *et al*.

Here, we employ the RT model to fit the temperature dependent relaxation time. The RT model was originally proposed to address the ultrafast relaxation mechanism in superconductivity. Later, it was shown to be suitable for a variety of materials with gap openings in the density of states. In these systems, photoexcitation produces large numbers of quasiparticles (QPs) that decay towards an initial equilibrium state through e-e or e-ph interactions. In RT model, the formation of an energy gap would introduce a phonon bottleneck effect, which would significantly impede the relaxation of QPs. The source of the phonon bottleneck effect is that high frequency phonons (HFPs) are generated when QPs recombine across the energy gap, which may in turn induce the excitation of a large number of QPs, thereby prolonging the relaxation time of the entire system. Based on this model, the density of thermally activated QPs $n_T$ can be obtained *via* the amplitude A(T), [42,43]

$$n_T = \frac{\eta}{4R}\left[\frac{A(0)}{A(T)} - 1\right] \qquad (4)$$

where η is the e-h pair regeneration rate due to annihilation of HFP, R is the recombination rate of e-h pairs, A(T) is the temperature dependent decay amplitude of the QP dynamics which we obtained by the single expotential fitting of transient THz transmission. The value of A(0) = 0.068 is determined by extrapolating the A(T) curve from T > 100 K to zero temperature [Fig. 3(c) solid line]. The blue dots in Fig. 3(c) are the fitting parameter A(T) with respect to temperature. Moreover, in the RT model, the decay rate of pump-probe dynamics could be described as: [44]

$$\tau^{-1} = \Gamma(\delta(n_T + 1)^{-1} + 2n_T) \tag{5}$$

Where $\Gamma = \frac{2R\gamma}{\eta^2(1+2\gamma/\eta)}$, R and η are defined above and γ is the decay rate of HFP. The coefficient Γ relates the QP density and the constant δ relates the photoexcited carrier density to the decay rate, both of them are independent of temperature. $n_T$ is given by the RT model, which takes a general form: [42,45]

$$n_T \propto T^{0.5} \exp\left(-\frac{\Delta}{2k_B T}\right) \tag{6}$$

where Δ is the gap energy. By fitting the temperature dependent decay time (Fig. 3(b)) and amplitude (Fig. 3(c)) simultaneously with the equations above, we obtain the temperature dependent $n_T$ as shown in Fig. 3(d), and the gap energy of Δ = 35 ± 6 meV, which is consistent with DFT-calculation [20] and ARPES measurement [46,47]. The deviations at temperature lower than 100 K in Figs. 3(c) and 3(d) may be related to the largely increase in mobility at low temperature. The red solid lines in Figs. 3 (b), (c) and (d) present the best fitting of temperature dependent A, τ, and $n_T$ with RT model. It is clear that the RT model can fit the experimental data well.

According to RT model, the increase in temperature results in more HFP being excited, which slows down the decay process as we see in Fig. 3(b) with temperature below 220 K. The increase in temperature also produces more thermal carriers $n_T$, leading to the increases of e-e scattering rate, and the e-h recombination rate could be accelerated assisted by hot carriers, as a result, the temperature dependent relaxation time is the balanced results between hot carrier promoted e-h recombination and HEP effect. Taking both the reflection and transmission of the film into account, the absorbed photon density is ~1.2×10$^{17}$ cm$^{-3}$ at 18.75 μJ/cm$^2$, and the photocarrier density reaches about 10% of the carrier density in equilibrium state. Therefore, the pump fluence-dependent decay time at room temperature hardly changes with the pump fluence as illustrated with blue dots in Fig. 3(a), where the slight decrease with pump fluence is due to the increase in hot carrier assisted e-h recombination rate. In contrast, the decay time is seen to increase with pump fluence at 5 K as shown in Fig. 3(a), which is believed to be caused by elevated temperature of the film with photoexcitation. According to the analysis above, photocarrier dynamics in $Cd_3As_2$ thin film conforms to the RT model, which reveals that the **Drude weight increases much more than the scattering rate after photoexcitation, resulting in the reduction of THz transmission. On the other hand,** the HFPs are generated across the narrow gap caused by the low dimensionality of $Cd_3As_2$ film, which can re-excite the recombined carriers, thus contributing a bottleneck effect to the recombination process.

To summarize, by employing temperature and pump fluence dependence of transient THz spectroscopy, we have investigated the photocarrier dynamics in a 50-nm $Cd_3As_2$ film. Photoexcitation leads to abrupt decrease of THz transmission, and the subsequent relaxation can be well reproduced with single exponential function. The temperature dependent relaxation time increase with temperature below 220 K, while the relaxation time remains ~7.5 ps with temperature above 220 K. The relaxation time shows pump fluence independence at 300 K, while it increases with pump fluence at low temperature. In addition, our pump fluence, delay time and temperature dependence of PC dispersion can be well fitted with pure Drude model, which verify that **Drude weight increases more than the scattering rate after photoexcitation, and the relaxation is dominated by depopulating carrier density *via* e-h recombination.** The RT model is employed to analyze the experimental findings, which demonstrate that gap opening with magnitude of Δ = 35 ± 6 meV occurs due to the reduced dimensionality in $Cd_3As_2$ film. The e-h recombination excites HFP across the narrow gap. In turn, the HFP can re-excited the carriers, which introduce a phonon bottleneck effect to slow down the carrier dynamics. Moreover, the fitting PC dispersion at various temperatures, pump fluence and delay times with Drude mode clearly demonstrate the applicability of the RT model, and further demonstrate that the photocarrier dynamics in $Cd_3As_2$ film follows a semiconductor-like behavior. These results are of deep insight into the whole carrier dynamics process in $Cd_3As_2$ and seeking potential applications in optoelectronics.

**ACKNOWLEDEMENTs**: This work is supported financially by the National Natural Science Foundation of China (NSFC, Nos. 92150101, 61735010).

## AUTHOR DECLARATIONS

**Conflict of Interest**

The authors have no conflicts to disclose.

## DATA AVAILABILITY

The data that support the findings of this study are available from the corresponding authors upon reasonable request.